\documentclass[aps,prl,10pt,twocolumn,superscriptaddress,balancelastpage,showpacs,reprint,english,floatfix]{revtex4-1}
\usepackage[T1]{fontenc}
\usepackage[latin9]{inputenc}
\usepackage{amsmath}
\usepackage{amssymb}
\usepackage{graphicx}
\usepackage{color}
\usepackage{soul}
\usepackage{hyperref}
\usepackage{graphicx}
\usepackage[export]{adjustbox}

\makeatother
\makeatletter

\newcommand{\qq}{\begin{eqnarray}}
\newcommand{\qqq}{\end{eqnarray}}

\newcommand{\p}{\partial}

\newcommand{\bfx}{\mathbf{x}}

\newcommand{\bfr}{\mathbf{r}}

\pdfpageheight\paperheight
\pdfpagewidth\paperwidth

\makeatother

\usepackage{babel}

\begin{document}
	
\title{Statistical properties of microphase and bubbly phase-separated active fluids}

\author{Giordano Fausti}
\affiliation{Service de Physique de l'Etat Condens\'e, CEA, CNRS Universit\'e Paris-Saclay, CEA-Saclay, 91191 Gif-sur-Yvette, France}
\affiliation{Max Planck Institute for Dynamics and Self-Organization (MPI-DS), 37077 G\"{o}ttingen, Germany}
\affiliation{Dipartimento di Fisica, Sapienza Universit\`a di Roma, P.le Aldo Moro 2, 00185 Rome, Italy}

\author{Michael E. Cates} 
\address{DAMTP, Centre for Mathematical Sciences, University of Cambridge, Wilberforce Road, Cambridge CB3 0WA, UK}

\author{Cesare Nardini}
\affiliation{Service de Physique de l'Etat Condens\'e, CEA, CNRS Universit\'e Paris-Saclay, CEA-Saclay, 91191 Gif-sur-Yvette, France}
\affiliation{Sorbonne Universit\'e, CNRS, Laboratoire de Physique Th\'eorique de la Mati\`ere Condens\'ee, LPTMC, F-75005 Paris, France}

\date{\today}
% ==============================================================================

\begin{abstract}
In phase-separated active fluids, the Ostwald process can go into reverse leading to 
either microphase separation or bubbly phase separation. We show that the latter is formed of two macroscopic regions that are occupied by the homogeneous fluid and by the microphase separated one. Within the microphase separated fluid, the relative rate of the Ostwald process, coalescence, and nucleation determines whether the size distribution of mesoscopic domains is narrowly peaked or displays a broad range of sizes before attaining a cutoff independent of system-size. Our results are obtained via large-scale simulations of a minimal field theory for active phase separation and reproduced by an effective model in which the degrees of freedom are the locations and sizes of the microphase-separated domains. 
\end{abstract}

\pacs{???}

\maketitle 

Active matter is a class of non-equilibrium systems in which each constituent particle consumes energy to produce work~\cite{ramaswamy2017active,Marchetti2013RMP}. Phase-separation is a common self-organization phenomena in active systems~\cite{Tailleur:08,Cates:15,Fily:12}. 
It was first described via an approximate mapping onto equilibrium liquid-vapor phase separation~\cite{Tailleur:08,Cates:15}, leading to speculation that detailed balance might be restored macroscopically in the steady state~\cite{Tailleur:08,Cates:15,Maggi:15,Brader:15,Speck2014PRL,fodor2016far,szamel2016theory,nardini2017entropy}. Indeed, perturbatively close to the liquid-vapor critical point, activity is an irrelevant perturbation~\cite{caballero2020stealth,caballero2018bulk,maggi2021universality,partridge2019critical} and nucleation processes can be partially explained by an extension of classical nucleation theory~\cite{cates2023classical,richard2016nucleation,redner2016classical,levis2017active}. 

It however became clear soon afterwards that active phase separation displays strongly non-equilibrium features. 
This is because activity has a crucial impact on interfacial tension~\cite{bialke2015negative,solon2018generalized2,tjhung2018cluster,fausti2021capillary,zakine2020surface,patch2018curvature,langford2023theory}, creating distinct tensions that describe the Ostwald process for liquid droplets and vapor bubbles~\cite{tjhung2018cluster}. An associated discovery was that the Ostwald process can go into reverse for one of the two phases, preventing the convergence at late times to bulk, equilibrium-like, phase separation~\cite{tjhung2018cluster}. Yet, capillary waves at the liquid-vapor interface are stable~\cite{fausti2021capillary,langford2023theory,patch2018curvature,lee2017interface}, so that the system remains phase-separated. (A regime of unstable capillary waves also exist and corresponds to a distinct phenomenology which we will address here~\cite{fausti2021capillary}.) In this regime active fluids are, depending on the global density, either microphase-separated ($\mu\textrm{PS}$) or bubbly phase-separated (BPS). In the former, $\mu\textrm{PS}$ domains retain a finite, system-independent, size in the steady state; in the latter, large liquid droplets contain a population of mesoscopic vapor bubbles that are continuously created in the bulk, coarsen, and are ejected into the exterior vapor. This creates a circulating phase-space current in the steady state, a feature that cannot arise in passive system~\cite{tjhung2018cluster}.

\begin{figure}[h!]
\includegraphics[width=1\linewidth]{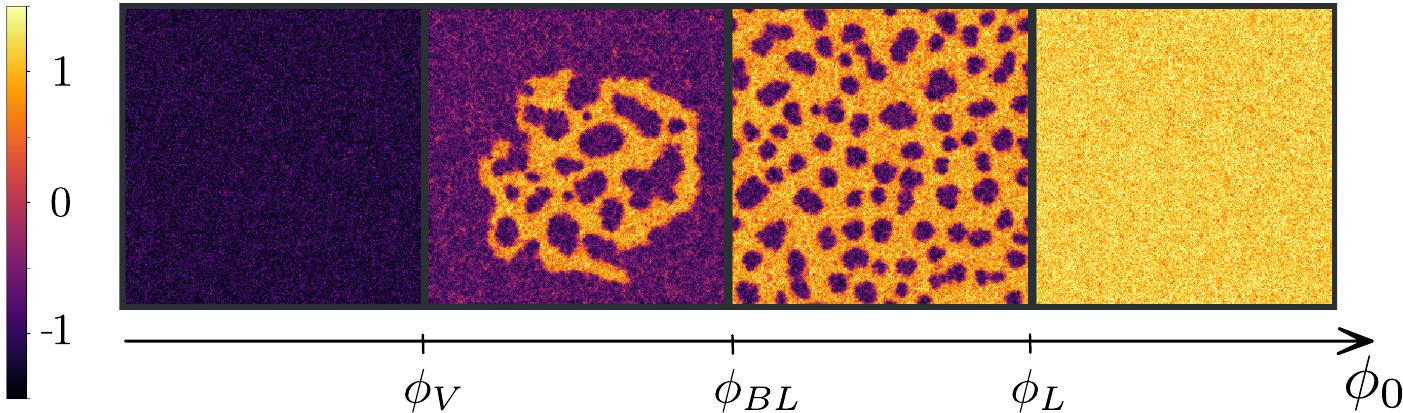}
	\caption{Phase diagram of AMB+ as a function of the global density $\phi_0$ in the region of reversed Ostwald process. From left to right: homogeneous vapor phase, bubbly phase separation, microphase-separation, and homogeneous liquid. Parameters used: $L_y=L_x=384$, global densities $\phi_0=-1.3,-0.4,0.4,1.3$.}
	\label{fig:snap_phasediag}
\end{figure}

Intriguingly, some of the simplest active matter models display a reversed Ostwald process, $\mu\textrm{PS}$ and BPS states. Self-propelled particles with hard-core repulsive interactions phase separate into dense (liquid) and dilute (vapor) phases because they slow down at high density by a mechanism known as Motility-Induced Phase-Separation~\cite{Tailleur:08,Cates:15,Fily:12}. 
A population of vapor bubbles in the dense phase is clearly visible in large-scale simulations both during coarsening~\cite{stenhammar2014phase} and in the steady state~\cite{caporusso2020micro,shi2020self,patch2018curvature,nakano2023universal}. More recently it was shown that the same models, when studied at high volume fraction, settle in a $\mu\textrm{PS}$ state formed of vapor bubbles~\cite{caporusso2020micro,shi2020self,nakano2023universal}. Vapor bubbles in the dense phase have been also seen in simulations of self-propelled particles that have hard-core repulsion and attract at larger distances~\cite{redner2013reentrant}, while a $\mu\textrm{PS}$ state was observed numerically and explained in terms of a reversed Ostwald process in the presence of intermittent attracting forces between particles, meant to mimic the pilii attachment performed in certain bacterial species~\cite{alston2022intermittent}. $\mu\textrm{PS}$ states are also commonly observed in experiments with self-propelled colloids~\cite{Palacci:12,Speck:13,van2019interrupted,thutupalli2017boundaries}, although system-specific explanations for some of these have been proposed~\cite{van2019interrupted,thutupalli2017boundaries,mani2015effect,matas2014hydrodynamic,mognetti2013living}.

Continuum descriptions of active systems have provided significant understanding of the generic, detail-independent, properties of active phase separation. The simplest setting consists in retaining only the evolution of the density field $\phi$~\cite{Wittkowski14,tjhung2018cluster,thomsen2021periodic}, while hydrodynamic fluid~\cite{tiribocchi2015active,singh2019hydrodynamically} or polar~\cite{tjhung2012spontaneous} fields can be added if the phenomenology requires. Their construction, via conservation laws and symmetry arguments, follows a path first introduced with Model B for passive phase separation~\cite{hohenberg1977theory,chaikin2000principles,bray2001interface}. (Recent works obtained them also via explicit coarse-graining procedures of specific particle models ~\cite{solon2018generalized,bickmann2020predictive,bickmann2020collective,kalz2023field,alston2022intermittent}.) Yet, these field theories differ from Model B because locally broken time-reversal symmetry implies that new non-linearities are allowed. The ensuing minimal theory, Active Model B+ (AMB+)~\cite{nardini2017entropy,tjhung2018cluster}, includes all terms that break detailed balance up to order $\mathcal{O}(\nabla^4\phi^2)$ in a gradient expansion~\cite{nardini2017entropy,tjhung2018cluster} and is defined by 
\qq
\p_t\phi&=&-\nabla\cdot\left(\mathbf{J}+\sqrt{2DM}\mathbf{\Lambda}\right)\,\label{eq:AMB+}\\
 {\bf J}/M &=&-\nabla \mu_\lambda  + \zeta (\nabla^2\phi)\nabla\phi\,\label{eq:AMB+J}\\
\mu_\lambda[\phi] &=& \frac{\delta \mathcal{F}}{\delta\phi} +\lambda|\nabla\phi|^2\label{eq:AMB+mu}
\qqq
where $\mathcal{F} = \int d\bfr \,\left[f(\phi) +\frac{K(\phi)}{2}|\nabla\phi|^2\right]$, $f(\phi)=a(-\phi^2/2+\phi^4/4)$ is a double-well local free energy,
and $\mathbf{\Lambda}$ is a vector of zero-mean, unit-variance, Gaussian white noises. Model B is recovered at vanishing activity
($\lambda=\zeta=0$), unit mobility ($M=1$) and constant noise level $D$~\cite{hohenberg1977theory}. It should be noticed that AMB+ is symmetric under $(\lambda,\zeta,\phi)\to-(\lambda,\zeta,\phi)$ so that all the conclusions below still hold exchanging the liquid with the vapor, so long as the activity parameters are also reversed in sign.

In this Letter we study the statistical properties of $\mu\textrm{PS}$ and BPS active fluids building on the suitability of a coarse-grained approach for studying their generic large-scale and long-time properties. We do so by performing large scale simulations of AMB+ and of a model defined in terms of the position and sizes of the $\mu\textrm{PS}$ domains (bubble model). We show that the BPS state is a bona-fide phase separated state whose properties are independent both of system-size and global density, and measure the critical density $\phi_{BL}$ marking the transition from the BPS to the $\mu\textrm{PS}$ states. We further show that vapor bubbles can either be almost monodisperse or broadly distributed as a power-law before attaining a system-size independent cut-off, and identify a key control parameter, governing the crossover between these regimes, as a ratio of mass transfers arising from Ostwald and coalescence processes.

Simulations of AMB+ were done with a parallel pseudo-spectral code employing Euler time-integration, $4/2$ dealiasing, in rectangular boxes of volume $V=L_x\times L_y$; we employed time step $dt=2\times 10^{-2}$ and checked that our results are robust by decreasing it. Unless specified otherwise, we use $4a=K=1$, and $\zeta=2\lambda=2$ so that the Ostwald process is reversed for vapor bubbles while being normal for liquid droplets; the liquid and vapor binodals are respectively given by $\phi_{1,2}=\pm 1$ for $D=0$. Fig.~\ref{fig:snap_phasediag} reports the phase diagram of AMB+ for $D=0.3$.
%Fig.~\ref{fig:snap_phasediag} reports the phase diagram for $D=0.3$. 
%In mean- for which the transition between normal and reversed Ostwald process happens at $\zeta=1$, and set $\zeta=2\lambda=2$ so that the Ostwald process is reversed 

%%%%%%%%%%%%%%%%%%%%%%%%%%%%%%%%
%%%%%%%%%%%%%%%%%%%%%%%%%%%%%%%%
%%%%%%%%%%%%%%%%%%%%%%%%%%%%%%%%
%%%%%%%%%%%%%%%%%%%%%%%%%%%%%%%%
We start our investigation from the $\mu\textrm{PS}$ state. Initialising the system in a homogeneous initial condition at density $\phi_0$, we first show that the time needed for reaching the steady state is independent of system size~\cite{supp}. 
Once this is attained, the probability distribution of the local density $P(\phi)$ has, as expected, a bimodal shape~\cite{supp}. Furthermore, for sufficiently high densities ($\phi_0> \phi_L$), the system is devoid of vapor bubbles and it is thus homogeneous instead of microphase separated. 
To measure $\phi_L$ we employed two procedures: first, we measured the number of vapor bubbles as a function of $\phi_0$ and identified $\phi_L$ as the density above which no vapor bubble is found. Second, we measured $\phi_L$ as the value at which the peak of $P(\phi)$ corresponding to the liquid density attains $\phi_0$. Both measures lead to $\phi_L\simeq 1.2$ for the parameters considered, and the result is unchanged by increasing system size~\cite{supp}.

\begin{figure}
     \centering
\includegraphics[width=1\linewidth,valign=t]{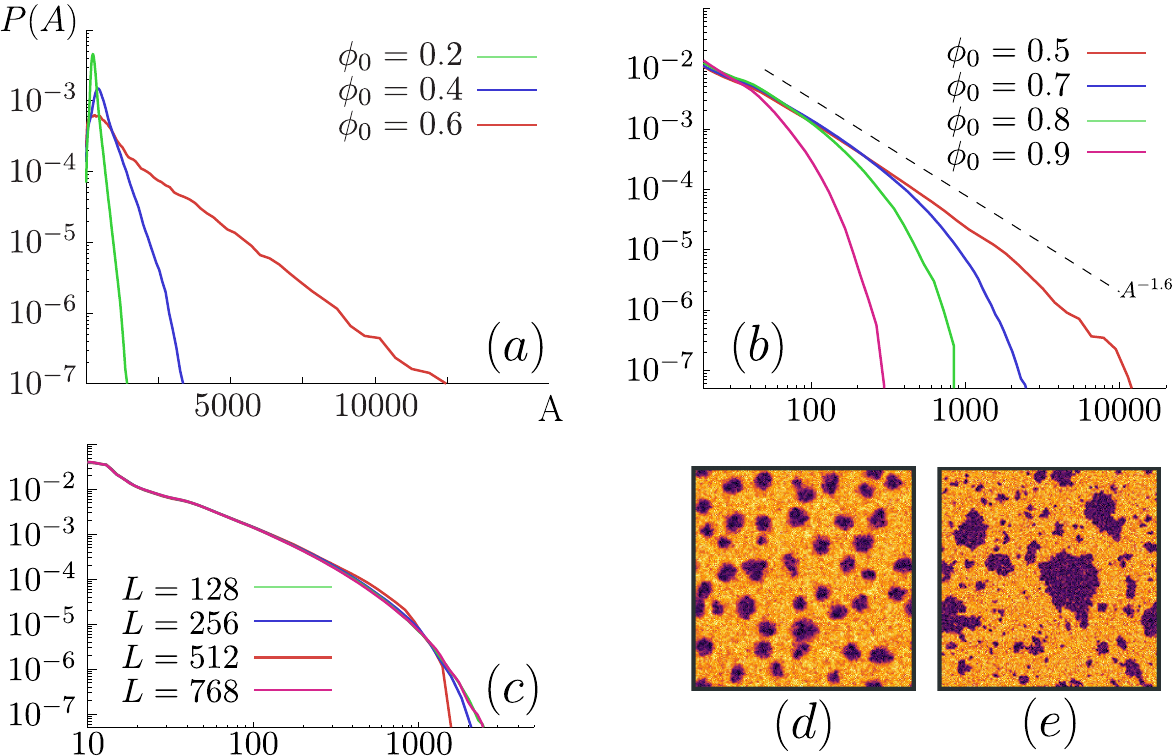}
    \caption{Distribution of the bubble sizes $P(A)$ in the $\mu\textrm{PS}$ state for various global densities $\phi_0$. This can be either narrowly peaked $(a)$ or broadly distributed $(b)$ displaying an intermediate power-law regime; notice that $(a)$ is in log-lin scale while $(b)$ is in log-log scale. (c) shows the system-size convergence of $P(A)$ for $\phi_0=0.7$ (other parameters as in (b)). Parameters used: (a)  $D=0.3,\zeta=2\lambda=2, 4a=K=1$; (b) $D=0.4, a=4K=1, \zeta=2\lambda=0.5$. $P(A)$ were measured in systems of size $ L_x=2L_y=512$ while snapshots are at $L_x=L_y=256$ with (d) $\phi_0=0.6$, (e) $\phi_0=0.5$. 
%    {\red (b) Fig is: all $L=256$ except for $\phi_0=0.5$ that is $L_x=L_y=512$. Sim at $512$ are running, we might have cleaner data}
}
	\label{fig:PDF-microphase-separation}. 
\end{figure}

Next, we measured the distribution, $P(A)$, of the bubble sizes or areas $A$, using the algorithm discussed in~\cite{fausti2021capillary}. We do so when varying the activity parameter $2\lambda=\zeta$, the global density $\phi_0$ and the noise strength $D$. 
Both the average bubble size $\langle A \rangle$ and their variance increases, either: i) approaching the transition between reversed and normal Ostwald process (located at $\zeta=2\lambda=1$ for the parameters considered \cite{tjhung2018cluster}); ii) decreasing $D$; or iii) decreasing the global density $\phi_0$ towards $\phi_{BL}$ so to approach the transition between $\mu\textrm{PS}$ and BPS states. The tails of $P(A)$ were found to be exponential, see Fig. \ref{fig:PDF-microphase-separation}(a) and~\cite{supp}. 
These observations are consistent with the expectation that the bubble size distribution is the more narrowly peaked the stronger the Ostwald process is with respect to coalescence and nucleation events. We will discuss this point later in more detail. 

\begin{figure}[h!]
	\centering
\includegraphics[width=1\linewidth]{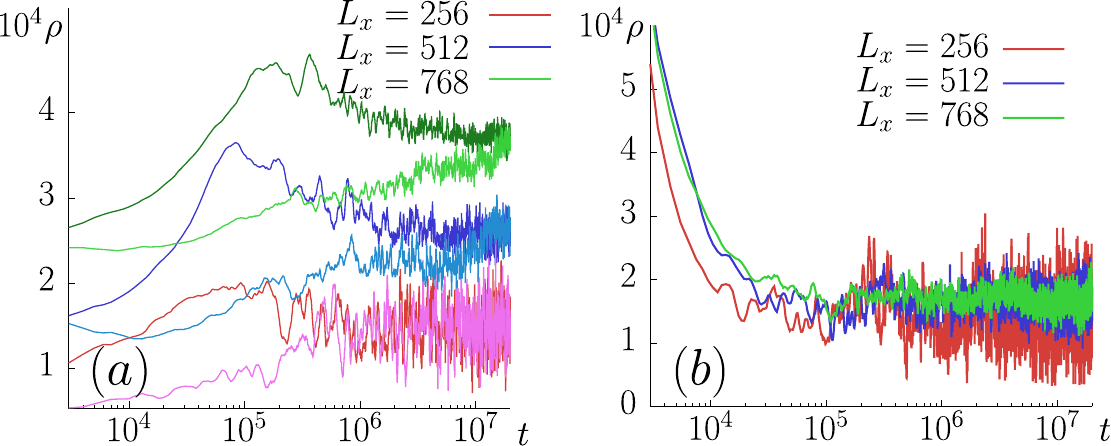}
	\caption{(a) Time-evolution of the number density of vapor bubbles $\rho$ starting from a bulk phase-separated state (dark color) and from a uniform initial condition (light color). The time scale needed to reach the steady state increases with $L_x$ in both cases. (b) Instead, when initialising the system from a phase-separated state in which the density of the dense region is set to the steady-state value $\phi_{BL}$, the relaxation time is independent of $L_x$. Parameters used $\phi_0=-0.2$, $L_x=2L_y.$ Time-series smoothed with a moving average with time-scale $5\times10^3$ (a) and
 $5\times10^4$ (b) and shifted vertically by $10^4$ ($L_x=512$) and $2\times 10^4$ ($L_x=768$) in (a) for visualization purposes.} 
	\label{fig:BPS-tauSS}
\end{figure}

The rate of the Ostwald process for vapor bubbles and the rate of relaxation of capillary waves are set by the associated interfacial tensions which, for $\zeta=2\lambda$, respectively read $\sigma=\sigma_{eq}(1-\zeta/K)$ and $\sigma_{cw}=\sigma_{eq}=\sqrt{8Ka/9}$. We now keep fixed the rate of the Ostwald process and of the relaxation of capillary waves and investigate how the $\mu\textrm{PS}$ state changes when varying the rate of nucleation and coalescence of vapor bubbles. To do so, we fix $\sigma$ and $\sigma_{cw}$ by keeping unchanged $\zeta/K=2$ and $Ka=1/4$ while varying $K/D$. At low $K/D$ more nucleation and coalescence events are present. Correspondingly, as shown in Fig. \ref{fig:PDF-microphase-separation}(e), the morphology of the bubbles now significantly deviates from a circular shape, and $P(A)$ displays a power-law regime $P(A)\sim A^{-\gamma}$, with $\gamma\simeq 1.6$. This power-law extends only for intermediate sizes, although on a large enough range to indicate an algebraic law; beyond a system-size independent cutoff, which increases when decreasing $\phi_0$, $P(A)$ decays exponentially. Intriguingly, in this regime, not only does the morphology of this $\mu\textrm{PS}$ state strongly resemble the one obtained in simulations of repulsive Active Brownian particles at high densities, but also the exponent of the power-law is similar ($\gamma\simeq 1.6$ in \cite{caporusso2020micro}). Movies in~\cite{supp} show the dynamics of the $\mu\textrm{PS}$ state in the regimes corresponding to Fig.~\ref{fig:PDF-microphase-separation}(a) and \ref{fig:PDF-microphase-separation}(b). 

%%%%%%%%%%%%%%%%%%%%%%%%%%%%%%%%
%%%%%%%%%%%%%%%%%%%%%%%%%%%%%%%%
%%%%%%%%%%%%%%%%%%%%%%%%%%%%%%%%
%%%%%%%%%%%%%%%%%%%%%%%%%%%%%%%%
\begin{figure}[!h]
\includegraphics[width=1\linewidth]{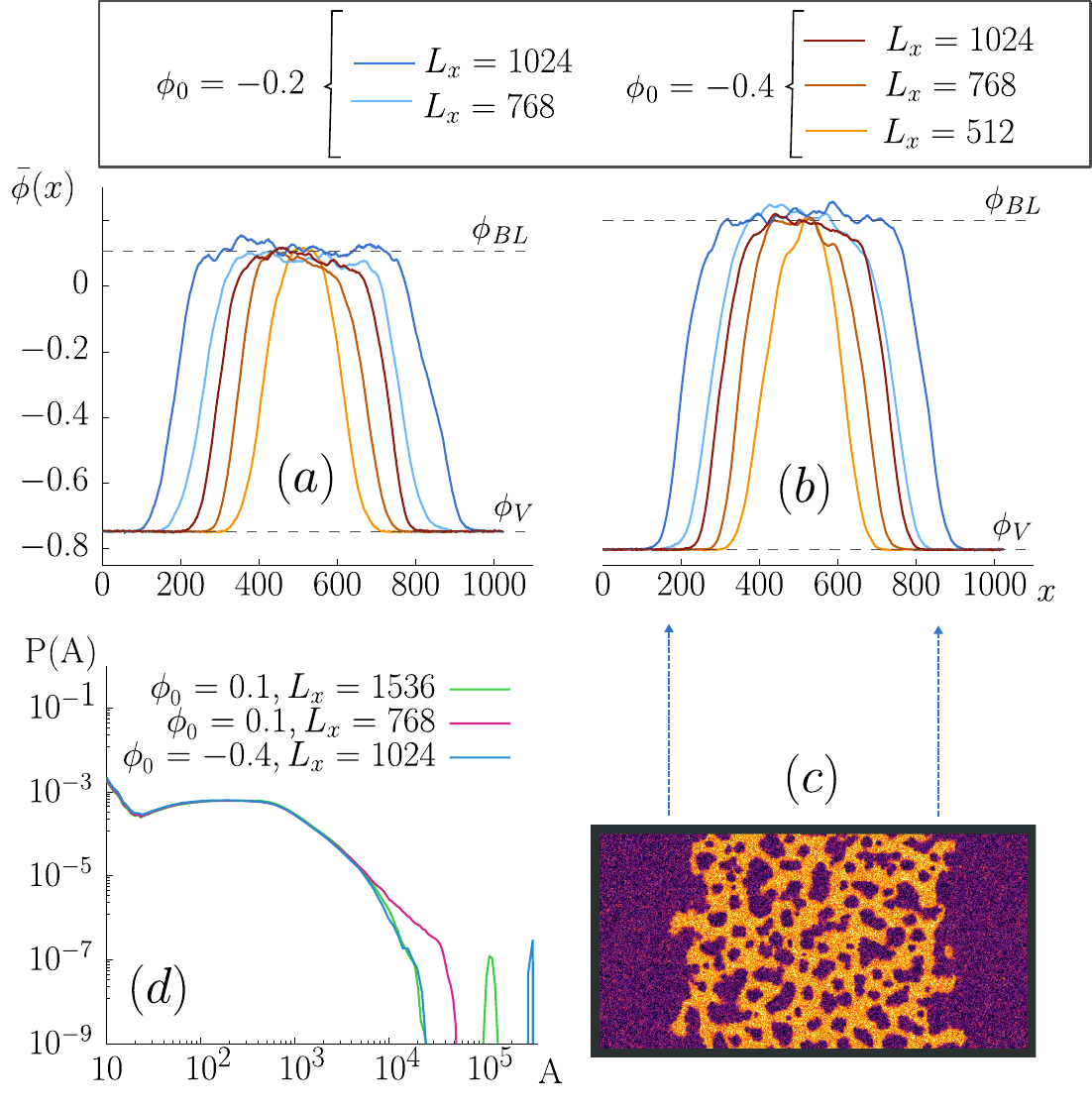}
 	\caption{Average density profile $\bar{\phi}(x)$ as a function of the $x$ coordinate for various global densities and system-sizes $L_x=2L_y$ when (a) $D=0.3$ and (b) $D=0.35$. The density of the vapor $\phi_V$ and the one of the $\mu\textrm{PS}$ state $\phi_{BL}$ can be clearly identified and are independent of both system size and $\phi_0$, as expected in a phase separation scenario. 
 (c) Snapshots of the BPS state for $\phi_0=-0.2$ and $L_x=2L_y=1024$.
 (d) When the global density is slightly below $\phi_{BL}$ (here, $\phi_0=0.1$ vs $\phi_{BL}=0.2$), the system looks $\mu\textrm{PS}$ at small system-size, but it becomes BPS asymptotically in system-size. Once bubbly phase-separation is attained, the distribution of vapor bubbles within the dense region is independent of global density $\phi_0$ and system size, and the two coexisting regions (filled of vapor and of the $\mu$PS state) are macroscopic (i.e., their size is proportional to the system-volume). }
	\label{fig:BPS-D0.25_profiles-main}
\end{figure}

We now turn our attention to $\phi_0<\phi_{BL}$ so that the system is expected to be BPS in the steady state. This is much more challenging computationally not only because lowering $\phi_0$ increases both the average bubble size and their variance, but also because the time needed to converge to bubbly phase-separation dramatically increases with system size. To show this latter fact, we performed simulations for different system sizes in a rectangular geometry starting either from a uniform initial condition (UIC) or from a band of liquid surrounded by vapor (BIC) initialised at the mean-field binodals $\phi=\pm 1$. Fig. \ref{fig:BPS-tauSS} shows the density of vapor bubbles as a function of time (other observables, such as the density of the $\mu\textrm{PS}$ region, give qualitatively similar behavior).

The slow convergence to bubbly phase separation can be rationalised as follows. To reach the stationary state starting from BIC, a macroscopic amount of vapor has to diffuse into the liquid phase to lower the overall density from the liquid binodal to $\phi_{BL}$. The diffusive vapor influx requires a time proportional to $L_x^2$, and our data indeed confirm this scaling~\cite{supp}. We now show that this vapor influx, and not the internal organisation of the $\mu\textrm{PS}$ state, is the crucial mechanism for the relaxation time being proportional to $L_x^2$. To do so, we modify the BIC by setting the density within the dense band to the steady state value $\phi_{BL}$: if in the original setting the relaxation time is dominated by the vapor influx across the interface, now the relaxation-time should be independent of $L_x$. As shown in Fig. \ref{fig:BPS-tauSS}(b) our simulations clearly confirm this expectation~\footnote{An additional slow relaxation mode due to the vapor density moving from the binodal to the steady state value might be present, but this is not visible in our data.}.

\begin{figure}[h!]\center
	\centering
\includegraphics[width=1\linewidth]{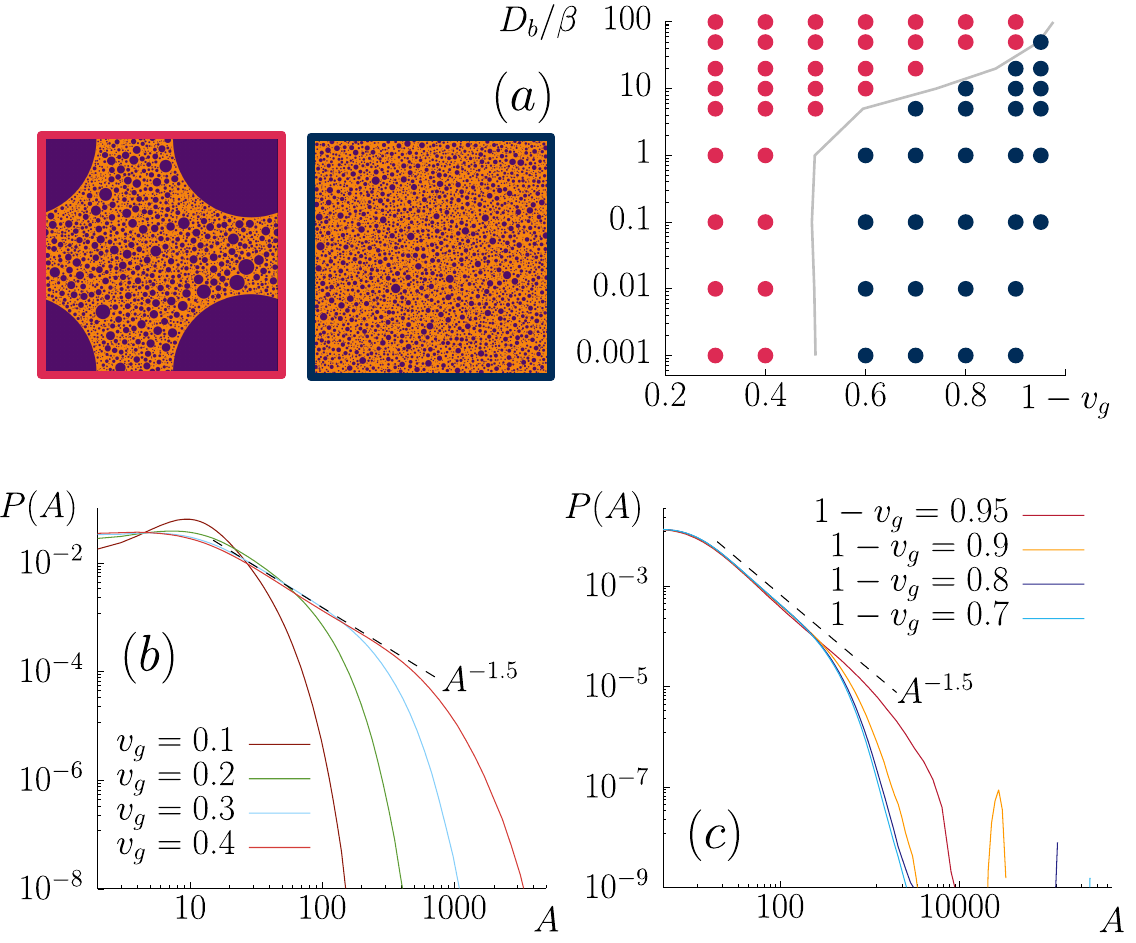}
	\caption{(a) Phase diagram of the bubble model with $\alpha=2$, $m_c=0.9$ for $L=800$. 
	For each $D_b/\beta$, we estimate the density separating $\mu\textrm{PS}$ from BPS regions (continuous line) by measuring the average density outside the largest bubble in simulations with $1-v_g=0.4$. Consistently, we observed either BPS (red dots) or $\mu\textrm{PS}$ states (blue dots) when performing simulations at other values of $v_g$. Interestingly, the transition density moves to higher values of $1-v_g$ when increasing  $D_b/\beta$. Spnapshots are at $D_b=5$,$v_g=0.6$ and $v_g=0.3$. (b) The variability of bubble sizes in the $\mu\textrm{PS}$ state increases when approaching the transition between $\mu\textrm{PS}$ and BPS and, when $D_b/\beta\simeq 1$ (here $D_b=5$), an intermediate power-law regime emerge in $P(A)$.	(c) When the global density is set in the BPS region ($D_b=10, v_g=0.4$ here), the system is $\mu\textrm{PS}$ for small $L$ and converges to BPS upon increasing $L$.}  
	\label{fig:EBM-phase-diagram}
\end{figure}

We next show that the BPS fluid is a bona-fide phase-separated state between two macroscopic regions: one filled of the $\mu\textrm{PS}$ state at density $\phi_{BL}$ and the other of the homogeneous vapor phase at density $\phi_V$. 
To show this we performed simulations in a rectangular geometry with $L_x=2L_y$ and considered the density field averaged over the $y$-direction and over time:
\qq\label{eq:bar-phi}
\bar\phi (x) =\frac{1}{(t_f-t_i) L_y}\int_{t_i}^{t_f}dt \int_0^{L_y} dy \,\phi(x,y,t)\,,
\qqq
where $t_i$ is the time at which the system reached the steady state.
Fig. \ref{fig:BPS-D0.25_profiles-main} shows the profiles of $\bar{\phi}$ as a function of $x$
for two noise levels. As expected in a phase separation scenario, $\bar{\phi}(x)$ plateaus to well-defined values, $\phi_{BL}$ and $\phi_V$, away from the interfaces, which depend neither on system size nor on the global density $\phi_0$. 
This measure, in turn, allows to define $\phi_{BL}$, and we obtained $\phi_{BL}(D=0.4)\simeq0$, $\phi_{BL}(D=0.35)\simeq 0.1$, $\phi_{BL}(D=0.3)\simeq 0.2$. 
To further probe the convergence in system size, we measured the full distribution of bubble sizes $P(A)$ and showed that it is independent both of size and global density~\cite{supp}. Our results are thus fully compatible with bubbly phase separation being a phase-separated state between the homogeneous vapor and the $\mu\textrm{PS}$ states at the minimal possible density $\phi_{BL}$, and provide a measure of $\phi_{BL}$ itself.

%%%%%%%%%%%%%%%%%%%%%%%

So far, we have studied the statistical properties of $\mu\textrm{PS}$ and BPS states at particular points in the phase diagram of AMB+. 
We now introduce a minimal bubble model, reminiscent of diffusion-limited aggregation models~\cite{krapivsky2010kinetic,majumdar2010real,majumdar2000nonequilibrium}, in which the degrees of freedom are the position and sizes of $N(t)$ vapor bubbles evolving in a square geometry of area $V=L^2$. Besides confirming the generality of the results already obtained within AMB+, we will obtain physical insight on the relative importance of the Ostwald process, coalescence and nucleation events for the statistical properties of the $\mu\textrm{PS}$ and BPS states. (The spontaneous evaporation of small vapor bubbles is also present at high noise but we neglect this in the following.) This is because, within the bubble model, their rates can be easily tuned while a similar exploration within AMB+ would require a computationally prohibitive exploration of the much larger parameter space.

We assume bubbles to be locally equilibrated, and thus circular, and describe them via the position of their center $\bfx_i$ and their radius $R_i>0$; we denote by $v_g$ the fraction of the volume they occupy (assuming the liquid and vapor densities to be $\pm 1$, $1-2v_g$ plays the role of the global density). Our bubble model is defined by four ingredients (see~\cite{supp} for details): {\it (i)} Bubble $i$ undergoes a reversed Ostwald exchange of mass with rate $\beta>0$ with its first nearest neighbour $\langle i \rangle$:
\begin{eqnarray}
		\dot R_i = \frac{\beta}{R_i}\left[ \frac{1}{R_i} -\frac{1}{R_{\langle i \rangle}} \right] \label{eq:Ost} \quad ;\quad 
		\dot R_{\langle i \rangle} = \frac{\beta}{R_{\langle i \rangle}}\left[ \frac{1}{R_{\langle i \rangle}} - \frac{1}{R_i} \right]\,.\label{eq:Ost_nn}
\end{eqnarray} 
{\it (ii)} Each bubble diffuses in space $\dot \bfx_i= \sqrt{2\tilde{D}_b/R_i^\alpha} \eta_i $, with a diffusion constant $\tilde{D}_b/R_i^\alpha$. 
{\it (iii)} Bubbles of radii $R_i\geq R_j$ coalesce upon touching and are replaced by a newly formed one located at their centre of mass with volume $\pi(R_i^2+m_c R_j^2)$, where $m_c\in(0,1)$ sets the amount of vapor lost in the coalescence event.
 {\it (iv)} Following each coalescence event we nucleate new vapor bubbles of radius $R_0$, whose total volume matches the one lost in coalescence (up to fluctuations of order $R_0^2$),
 and randomly place them.

A few comments are needed before presenting our results. In an earlier attempt at building a bubble model~\cite{shi2020self}, the Ostwald process was represented as a deterministic shrinking of all bubbles, its $R$-dependence was not realistically reproduced, and the diffusion constant of vapor bubbles was set to be independent of its volume (corresponding, in our notations, to $\alpha=0$). Instead, our eq. (\ref{eq:Ost_nn}) was obtained as the dynamics of a pair of isolated bubbles~\cite{tjhung2018cluster}, and thus more faithfully represents the Ostwald process in a minimal fashion. Furthermore, the diffusion constant of vapor bubbles with large radii is known to be inversely proportional to their volume if momentum is not conserved. The result was first shown for Model B~\cite{Onuki,kawasaki1983kinetics} and gives $\alpha=2$ and $\tilde{D}_b\propto D$. This is reproduced in our bubble model by fixing $\alpha=2$, and we make this natural choice below. Because vapor bubbles are mesoscopic in size, one might however wonder if imposing $\alpha=2$ is crucial for our results; we thus show that this is not the case and very similar results are obtained using $\alpha=3/2$ and $\alpha=1$ albeit with stronger finite-size effects when decreasing $\alpha$~\cite{supp}. Third, we have chosen $m_c<1$, corresponding to assuming that vapor is lost during coalescence events. This is however natural in the regime in which bubbles are far apart and equilibrated to the local liquid environment: as shown in~\cite{supp}, when two bubbles of radius $R$ coalesce, the newly formed one will shrink (back to $R$ in the limit in which other bubbles are infinitely far away). It should be observed that this does not require the interaction with distant bubbles via the reversed Ostwald process, as it happens also for two isolated bubbles in a domain much larger than their radius. Because the newly formed bubble shrinks to equilibrate to the local liquid environment, the density of the liquid surrounding decreases, and hence the nucleation probability increases. Our steps $(iii)$ and $(iv)$ above take these processes into account in a minimal way. 

We set in the following $R_0=1/2$ and $\beta=1$, thereby fixing the units of length and time. Three free parameters are left: the total density (set by the volume occupied by the vapor $v_g$), $m_c$, and the non-dimensional ratio $D_b/\beta=\tilde{D}_b (2 R_0)^{1-\alpha}/\beta$, measuring the relative strength of coalescence {\emph versus} the Ostwald process.  $D_b/\beta$ crucially determines the statistical properties of the system; Fig. \ref{fig:EBM-phase-diagram}(a) reports its phase diagram as a function of the volume fraction occupied by the liquid $1-v_g$ and of $D_b/\beta$ for $m_c=0.9$. As in AMB+: the system is $\mu\textrm{PS}$ at high density (blue dots) and BPS (red dots) at low density ($v_g>v_{BL}$). The continuous line is $1-v_{BL}$ as obtained from measuring the density in the $\mu\textrm{PS}$ region when the system is BPS: as expected, this accurately locates the transition between $\mu\textrm{PS}$ and BPS states. Convergence in system size is achieved, and discussed in~\cite{supp}. 

The bubble model reproduces most of the properties already described within AMB+~\cite{supp}: while the convergence in time to the $\mu\textrm{PS}$ state is independent of system size, it increases with it when converging to the BPS state. 
Second, the $P(A)$ for the $\mu\textrm{PS}$ state is peaked and has exponential tails when either $D_b/\beta\ll 1$ or  $D_b/\beta\gg 1$, and the bubble size variability increases when $v_g$ approaches the transition between $\mu\textrm{PS}$ and BPS states at $v=v_{BL}$. Third, when the rate of mass redistribution due to the Ostwald process and to coalescence are comparable ($D_b/\beta\simeq 1$), $P(A)$ displays an intermediate power-law regime at intermediate values of $A$. Such a power-law regime extends for a broader range of sizes when decreasing $\alpha$. Finally, setting $v_g\gtrsim v_{BL}$, the system is $\mu\textrm{PS}$ for small system-size, but converges to the BPS state increasing $L$. Furthermore, the bubble model predicts that the transition from $\mu\textrm{PS}$ to BPS is located at $v_{BL}\simeq 1/2$ whenever the Ostwald process is much faster than mass redistribution due to coalescence and nucleation ($D_b/\beta\ll 1$) see Fig. \ref{fig:EBM-phase-diagram}(a) and that $v_{BL}$, the density separating $\mu\textrm{PS}$ from BPS states, increases towards the liquid binodal upon increasing $D_b/\beta$. This latter fact is a new prediction of the bubble model, which at this stage is hard to check in AMB+ as it is unclear how to vary the coalescence and nucleation rates in terms of the AMB+ parameters without changing other properties.

%%%%%%%%%%%%%%%%%%%%%%%%%%%%%%%%
%%%%%%%%%%%%%%%%%%%%%%%%%%%%%%%%
%%%%%%%%%%%%%%%%%%%%%%%%%%%%%%%%
%%%%%%%%%%%%%%%%%%%%%%%%%%%%%%%%
In summary, we studied the statistical properties of phase-separating active fluids subject to a reversed Ostwald process. We did so with two complementary approaches: large-scale simulations of AMB+, and a minimal model in which the degrees of freedom are the vapor bubbles. Our results show that microphase separated domains have, in different regimes, sizes that are either narrowly peaked, or distributed over a broad range of scales. The latter echoes previous finding in models of self-propelled particles, in which vapor bubbles were observed to be broadly distributed~\cite{shi2020self,caporusso2020micro,nakano2023universal}. Furthermore, we have shown that bubbly phase separation is a bona-fide phase-separated state in which both the homogeneous and the microphase-separated regions are macroscopic in size. 
It is currently unknown whether one can tune microscopic parameters in systems of self-propelled particles to get microphase separated active fluids formed of mono-disperse domains. It is likewise unknown even in the simplest particle models how $\phi_{BL}$, the density separating bubbly from microphase separation, depends on them. These questions merit further investigation. %Finally, our results might have relevance for other phase separating systems with locally broken detailed balance, from describing biological condensates~\cite{banani2017,Weber2019review} to sociophysics~\cite{grauwin2009competition,zakine2023socioeconomic}. 

% ==================================================================
% ==================================================================
\begin{acknowledgments}
The authors acknowledge A. Celani, H. Chat\'e, A. Patelli, J. Stenhammar for discussions and allocation of CPU time on the Living Matter Department cluster in MPIDS. GF was supported by the CEA NUMERICS program, which has received funding from the European Union's Horizon 2020 research and innovation program under the Marie Sklodowska-Curie grant agreement No 800945. CN and MEC acknowledge the support of 
the Institute National de Physique (INP) through the International Research Program (IRP) ``IFAM''. Work funded in part by the ANR grant ``PSAM'' and by the European Research Council under the Horizon 2020 Programme, ERC grant agreement number 740269 and by the National Science Foundation under Grant No. NSF PHY-1748958, NIH Grant No. R25GM067110. The authors would like to thank the Isaac Newton Insitute for Mathematical Sciences, Cambridge, for support and hospitality during the program ``Anti-diffusive dynamics: from sub-cellular to astrophysical scales'' where work on this paper was undertaken. This work was supported by EPSRC Grant No EP/R014604/1. This work was partially supported by a grant from the Simons Fundation. 
\end{acknowledgments}
% ==============================================================================

\bibliographystyle{apsrev4-1}
\bibliography{biblio.bib}

\end{document}